# Model Callers for Transforming Predictive and Generative AI Applications

**Mukesh Dalal**
AIDAA Services LLC
Metuchen, NJ 08840, USA
mukesh@aidaa.ai[1]

## Abstract

We introduce a novel software abstraction termed "model caller," acting as an intermediary for AI and ML model calling, advocating its transformative utility beyond existing model-serving frameworks. This abstraction offers multiple advantages: enhanced accuracy and reduced latency in model predictions, superior monitoring and observability of models, more streamlined AI system architectures, simplified AI development and management processes, and improved collaboration and accountability across AI/ML/Data Science, software, data, and operations teams. Model callers are valuable for both creators and users of models within both predictive and generative AI applications. Additionally, we have developed and released a prototype Python library for model callers, accessible for installation via pip or for download from GitHub.

## 1 Introduction

Since artificial intelligence (AI) systems[2] are increasingly being used across business and consumer applications, many new platforms and tools have been developed, especially for training[3] AI models and serving them at scale[4]. Our primary focus is on the other side of model serving, that is, how to improve the calling of AI models using a novel software (SW) abstraction called "model callers.*"* However, we will also show that model callers benefit both sides, that is, model serving and model calling. In particular, we propose that model callers could be the right abstraction for developing, upgrading to, and using compound AI systems [1] for SW 2.0 [2], while reducing technical debt [3, 4] and improving quality [5], tooling [6], operations [7], and lifecycle management [8].

### 1.1 Motivating use cases

We will first motivate the need for model callers with some practical use cases, before getting into the details of our approach and its benefits. We will build our use cases from a common base scenario in which we run a financial credit card processing system that relies critically on a fraud-detection component (FDC) for identifying fraudulent transactions [9] in our business.

---

[1] We highly value your constructive feedback and insights to enhance our approach and paper, especially as this is our initial public draft. Please feel free to share your thoughts via email or connect with us on Twitter/X @mukdal. Your input is crucial to our improvement process!

[2] We adopt a broad definition of artificial intelligence (AI), encompassing machine learning (ML), data science (DS), statistics, and related fields. Similarly, we use the term 'systems' in a comprehensive manner to refer to applications, apps, services, solutions, and other related entities.

[3] We use the term 'training' in its broadest sense, encompassing various methodologies such as pretraining, finetuning, instruction tuning, alignment tuning, incremental training, partial training, fitting, and partial fitting, among others.

[4] For example, Amazon SageMaker, databricks.com and neptune.ai.

### 1.1.1 AI transformation

Suppose FDC is based on the old rules-based technology [10] causing many problems, so we would like to upgrade it to use AI models. However, we have certain operational requirements, including: the system can't be taken out of operations for an upgrade because it gets used every second of the day, and the fraud-detection accuracy can't drop below the current level because that would cause significant costs and losses to our business. Our AI experts have identified some specific AI models (or vendors) that have demonstrated excellent accuracy results in experimental settings but can't ensure that these models will perform well in our actual operations and will continue to adapt well in the future. How should we resolve these concerns, since not upgrading is also not an option?

### 1.1.2 AI observability and accountability

Suppose we have somehow successfully upgraded FDC to use the new AI models. Our AI experts are observing 99.99% accuracy in their experiments, consistent with what the vendor is claiming across all customers, but our European business has anecdotal evidence of much worse accuracy. We ask our SW and operations (Ops) experts to validate accuracy for European transactions, but they are unable to do so without a massive upgrade to the data collection infrastructure which was designed to support only global. Could we have designed our overall AI system so that our SW and Ops teams could independently observe and validate performance of AI models at operational (regional) granularity, instead of just relying on the global observability reported by the AI team?

### 1.1.3 Granular fine-tuning of AI models

Suppose we have somehow validated claims from our European business that they are getting only 98.7% accuracy from FDC, while our American business is getting a much better 99.70% accuracy. We ask our AI team to further optimize the accuracy of their models for the European transactions, but they are unable to do it without a massive upgrade to their model training and serving infrastructure, which currently optimizes the model only at the global level. Could we have designed our system such that models get fine-tuned at regional levels, instead of being trained and tuned just at the global level?

### 1.1.4 Granular custom models

Suppose we have somehow fine-tuned the models deployed for the European business differently than for the American business and improved their regional accuracies to 99.80% and 99.99% respectively. One of our trusted vendors has pitched a new model that seems to work very well for the European transactions but not for the American transactions. However, serving a completely different model for Europe requires another massive upgrade in our infrastructure. Could we have designed our system such that completely different models are served at granular operational levels?[5]

### 1.1.5 Quickly onboarding new models

Suppose our system somehow enables serving a completely new model for Europe. However, we still need to onboard (train and/or fine-tune) this model for our specific operations. How should we get reliable training data needed for this? Note that synthetic or 3rd-party data may not reliably capture our operational data patterns.

### 1.1.6 Ensembles of models, etc.

---

[5] There's a strong theoretical backing for this use case, since it follows from the well-known no free lunch theorems [11] that there is no single optimization (i.e., model) for any non-trivial task that outperforms all other models for all data patterns.

Suppose we have somehow identified multiple good models for Europe. Can we deploy them all in an ensemble so that the final prediction is obtained by some aggregation of their individual predictions? Using an ensemble of models also increases the resiliency of our system since it will continue to work even if some of these models crash. Going further, suppose someone has programmed a traditional non-AI SW 1.0 function[6] that provides high accuracy for some specific data patterns. How should we include this non-AI function in our ensemble, just like the AI models? Our approach should work for the various ways such functions could be available to us, for example, provided by a vendor as a cloud service packaged in a Docker container.

### 1.1.7 Model and data drift

Suppose we have somehow properly trained multiple ensembles and deployed them in production. However, six months have gone by, and the data patterns have changed (even the models may have changed if they have been updated offline, say, by the vendor). How should we automatically identify whether the accuracy has dropped below acceptable levels at some regional levels?

### 1.1.8 Development and operational ease and automation

Even if we have somehow managed to cover all the above use cases, how should we make all of this super-easy for our teams, and even automate most of it, so that we avoid manual errors, improve performance, and decrease our total cost of ownership?

### 1.1.9 Future-proofing AI systems

How should we design our AI system such that it can be easily adapted to emerging and future technologies, such as more advanced language, vision and other foundation models as well as increasingly powerful frameworks for integrating them with external information, tools and systems?

## 1.2 Introduction to model callers

We will now present an overview of our approach for building AI systems based on model callers. Figure *1* illustrates two common model-calling patterns in current AI systems. In System A, AI models are called directly by other system components (indicated by empty boxes). In System B,

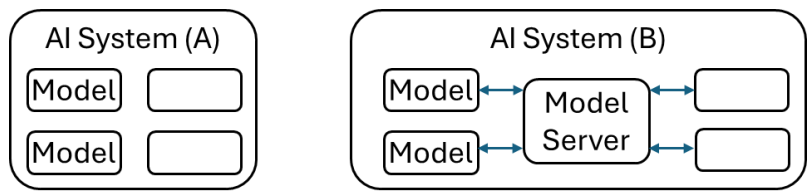

Figure 1: Two model-calling patterns in current AI systems. In System A, AI models are called directly by system components (indicated by empty boxes). In System B, AI models are accessed through serving-side intermediaries such as a model server.

AI models are accessed through serving-side intermediaries such as a model server. There are many variants of System B, such as encapsulating the models and the model server in a prediction service exposed through a REST API endpoint. The model server may provide or support sophisticated capabilities such as observability, management, and governance, often as part of MLOps, AIOps, LLMOps, etc. These capabilities are typically deployed on the model serving side, instead of the calling side.

---

[6] We use the term "function" in its broadest sense that includes methods, programs, services exposed thorough functional APIs, and even non-deterministic functions that are not mathematical functions.

Figure *2* illustrates how these patterns may be extended by inserting model caller intermediaries on the calling side. Each model call goes through a model caller which may (as in B) or may-not (as in A) use a model server to access the models. Technically, each model caller allows *registering* models which can then be accessed through it. By explicitly adding model callers in the model-calling patterns, we sharpen the clear distinction between the two sides: a *model-calling* side (or just the calling side) and a *model-serving* side (or just the model side or the serving side). These model caller intermediaries may be implemented in many ways, such as objects in a programming language like python, SW components, or as Docker-containerized services.

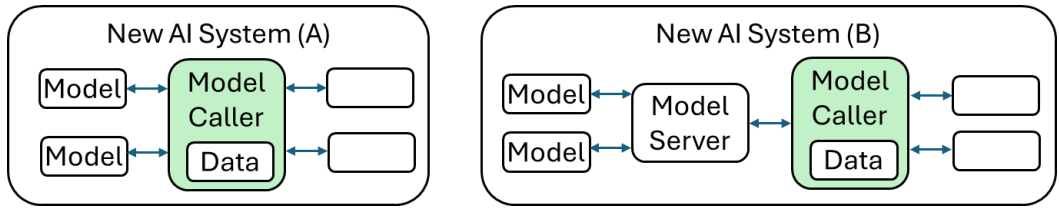

Figure 2: New AI systems extend the model-calling patterns of Figure 1 by inserting calling-side intermediaries called model callers (shaded). Model callers may cache call and other kinds of data.

Model callers may provide a variety of capabilities on the calling side, including the following (Section 2 provides more details):
- Automatically adding additional data from the calling side that could be used on the serving side of the model calls in multiple ways, such as traditional features in model training and inferencing, or as "hyper-features" for influencing model training and inferencing in other ways.
- Typical serving-side capabilities such as observability, management, governance, etc. which are not being provided or are only weakly provided by the current infrastructure.
- Duplication of serving-side capabilities on the calling-side for purposes such providing independent validation or providing additional granular information.
- Completely new capabilities that may be best provided on the calling side, such as call encryption and automated calling.
- Simplifying the architecture and lifecycle of AI systems by providing a better home, design, or evolution of existing, improved, or new capabilities.

While these model callers may be initially deployed on the calling-side, sometimes as prototypes or experiments, some of them may eventually get transitioned to the serving side. Even then, more model callers may be deployed on the calling side for calling these model callers on the serving side, and these model caller chains may get longer and more complex over time. In general, a model caller provides all the functionality of models with much more visibility and control, especially on the calling side.

### 1.3   Our contributions

Our core contributions in this paper include:
- Introduction of a new SW abstraction called model caller for improving AI model use in AI systems.
- Demonstrating the use of model callers for improving existing capabilities and delivering new capabilities for AI systems.
- A new python library, called `modelcaller`, that provides model caller capabilities out of the box through a new python class called `ModelCaller`.[7]
- Several specific examples of model caller capabilities and use cases.

---

[7] This library may be downloaded from https://github.com/mukdal/modelcaller for non-commercial use. Please contact us for limited free commercial licensing.

*A guide for the rest of the paper:* Section 2 describes various kinds of model callers and their capabilities. Section 3 is an initial unifying framework for model callers. Section 4 presents more details about the python library. Section 5 covers related work not covered earlier. Section 6 presents some conclusions.

## 2   Model callers: kinds and capabilities

We describe various kinds of model callers and their capabilities, starting with simple ones. Since these kinds overlap, a specific model caller may be of multiple kinds exhibiting a combination of their capabilities.

### 2.1   BasicMC

A basic model caller, called basicMC, registers exactly one AI model. Calling a basicMC results in invoking this registered model, with the following additional steps:
- The inputs and outputs of each call are automatically saved by the basicMC.
- The saved data is automatically partitioned into two categories, called *training* and *evaluation*, based on a configurable parameter that specifies the desired fraction of evaluation data.[8]

These basic capabilities are sufficient to provide the regional observability needed for the use case of Section 1.1.2. In particular, one may create two basicMCs, one for the European transactions and the other for the American transactions (additional ones may be created to support other regions), both registering the same model, as illustrated in Figure 3 where MC-1 and MC-2 are the two basicMCs. Region-specific metrics, such as model accuracy for European transactions, may then be computed on the calling side using the data saved in these basicMCs.

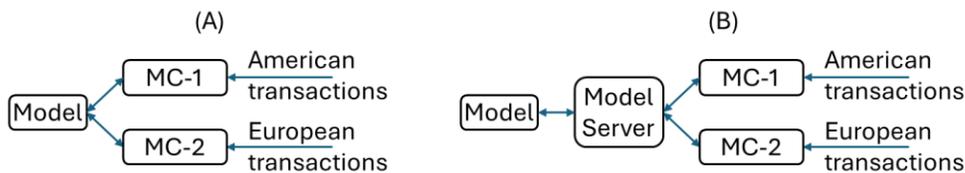

Figure 3: Two BasicMCs (MC-1 and MC-2) registering and accessing the same model to provide granular calling-side observability (for patterns (A) and (B) of Figure 2).

BasicMCs (even without any data saving) are also sufficient for the granular custom model use case of Section 1.1.4. In particular, one may again create a different basicMC for each region and register a different custom model with each of them.

Besides providing observability over the saved data, a basicMC may also provide management and governance capabilities over that data, such as compressing, archiving, encrypting, auditing, logging, and filtering of data.

### 2.2   SupervisoryMC

A supervisoryMC extends a basicMC by adding a supervisory interface that enables a user to review some outputs to selectively confirm or override them, as shown on the left side in Figure *4*. To control the supervisor load, it uses a configurable parameter to specify the fraction of outputs that are randomly selected for supervision. The right side in the figure illustrates a simple standalone GUI (graphical user interface) provided by a supervisoryMC. This GUI may also be integrated into the supervisory GUI for the entire AI system.

---

[8] Variants of basicMC may provide additional categories of data, such as *validation*, and use other methods for splitting the input/output data into these categories. For brevity and clarity, we will no longer explicitly mention many possible variants since they can be deduced by a careful reader.

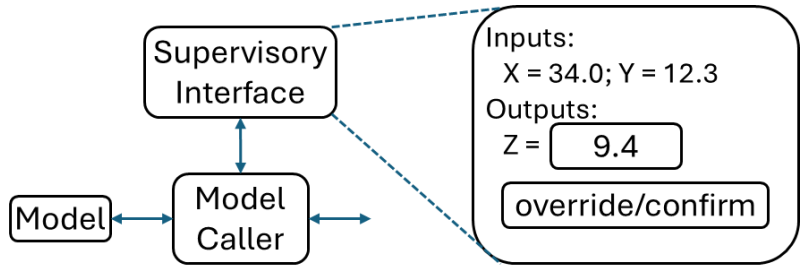

Figure 4: SupervisoryMC calling pattern on the left and a sample GUI on the right, corresponding to System A of Figure 2. System B is similar and thus omitted.

A supervisoryMC orthogonally partitions saved input/output data into two more categories: *supervised* (where a user has confirmed the output) and *unsupervised* (otherwise). This leads to four sub-categories of saved data, as illustrated in Table *1* along with their approximate relative sizes:
- Gold data: high quality and high importance, since it is both supervised and used for model evaluation (which we suppose to be more important than training)
- Platinum data: high quality, since it is supervised data
- Silver data: high importance, since it is used for model evaluation
- Bronze data: may not be high quality or importance

Table 1: Categories of data saved in a supervisoryMC

| Data categories | Supervised | Unsupervised |
|---|---|---|
| Evaluation | Gold (smallest) | Silver (large) |
| Training | Platinum (small) | Bronze (largest) |

SupervisoryMCs enable sampling and estimating model quality[9] in operational settings post deployment. Thus, they can help detect model and data drifts needed for the use case of Section 1.1.7.

## 2.3 EnsembleMC

An ensembleMC extends a basicMC by registering additional models, effectively creating a traditional ensemble model[10] enhanced with model caller capabilities. Thus, they provide standard benefits of ensemble learning, such as:
- Improved Accuracy: Combines predictions from multiple models to reduce variance and bias, often resulting in higher accuracy than any single model.
- Reduced Overfitting: Diverse models in the ensemble can average out individual errors, decreasing the likelihood of overfitting to the training data.
- Increased Robustness: Ensembles are more resilient to noise and outliers in the data, providing more stable predictions across varied datasets.
- Handling of High Dimensionality: Ensemble methods can better manage high-dimensional spaces and complex data structures.
- Model Uncertainty Estimation: Some ensemble techniques can provide measures of uncertainty or confidence in the predictions.
- Flexibility in Model Choice: Ensembles can combine different kinds of models, leveraging the strengths of each.
- Enhanced Generalization: By aggregating diverse models, ensembles can generalize better to unseen data, improving predictive performance.

---

[9] We use the term quality in its most general sense that includes accuracy, adequacy, etc.
[10] See [12] for a recent survey on ensemble learning.

- Error Correction: Techniques like boosting sequentially correct errors made by previous models, continuously improving performance.
- Parallelization: Many ensemble methods, such as bagging, can train models in parallel, making efficient use of computational resources.
- Compositions: Simpler systems (subsystems, logical components, agents) are composed to create more complex systems (supersystems).

An ensembleMC has the same calling pattern as the model callers in Figure *2*. Some of the models in an ensemble may be obtained from a single base model, as in snapshot ensembling (different training snapshots of the same base model) and neural weight diversification (different initializations and/or hyperparameters of the same base model).

Since there are two general kinds of ensembles (based on their overall control strategy), called *parallel* (bagging, stacking, voting, mixture of experts, snapshot ensembling, neural weight diversification, etc.) and *sequential* (boosting, random forest, etc.), we get two kinds of ensembleMCs, as described in the next two sub-sections.

### 2.3.1 ParallelMC

A parallelMC calls each registered model in parallel with the same inputs and then aggregates their individual outputs to obtain the final output of the model call. A parallelMC generally requires that all its registered models must have the same input and output parameters. There are many ways to aggregate the outputs, such as:
- Stacking: outputs from all models are just stacked together to produce a larger output.
- Voting (for categorical outputs): output the category with the most votes.
- Simple means: average across all models for each output parameter.
- Quality-weighted means: weigh each model's output with some measure of its estimated (or historic) quality.
- Aggregation model: use a custom AI model for aggregation.[11]

A parallelMC may also split its saved data among its individual models, similar to bagging.

### 2.3.2 SequentialMC

A sequentialMC calls its models in a sequence, where the outputs from earlier models may be used in the inputs for the later models, similar to boosting.

## 2.4 IdMC

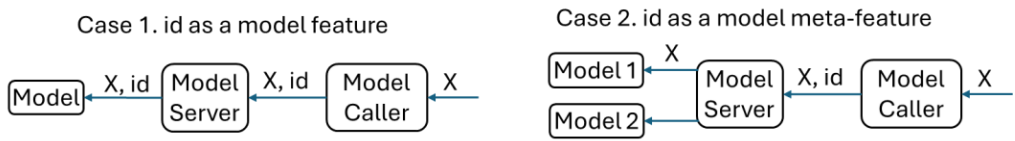

Figure 5: IdMC calling patterns where the model caller inserts its id as a call argument. In Case 1, the model server passes the id to the model as a feature. In Case 2, the model server uses the id as a "meta-feature" along with other arguments X to decide which model to invoke (Model 1 for this call).

An idMC extends a basicMC by automatically inserting its unique id as an additional argument before calling the registered model. This id could be anything that differentiates among different model callers. The direct serving-side partners of idMC (either the models or model-servers) must support this additional parameter, as illustrated in Figure *5*. This id may be passed to the model as a feature (Case 1) or may-be used by a model-server as a "meta-feature", say, for deciding which model to invoke (Case 2).

---

[11] This aggregation model has different input and output parameters than the models in the ensemble. It may even handle ensemble of models that differ in their output parameters.

An idMC enables using the same basicMC for different regions in supporting many of the use cases of Section 1.1, since the id allows the serving side to differentiate among those calls.

## 2.5 SensorMC

A sensorMC extends a basicMC by allowing sensors to add additional data without even calling the registered model. A sensorMC creates additional methods that may be used for adding input/output data from indirect sources (not direct calls to the model caller), as shown in Figure *6*. The data saved in a sensorMC is further partitioned depending on whether it was obtained from a model call or a sensor call.

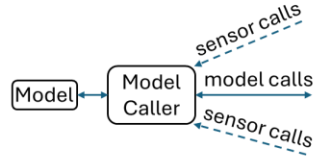

Figure 6: A sensorMC enables sensor calls just for adding data, in addition to regular model calls.

To motivate this, consider a sensorMC that registers a model which approximates a function `fn` such that its inverse function `finv` can be well-approximated by a hand-coded Java function. Now anytime the model is called via sensorMC, its input and output will be saved. However, additional input/output data may be generated by calling `finv` on randomly generated inputs and then calling the model caller's sensor method after reversing the inputs and outputs.

## 2.6 ReviewerMC

A reviewerMC extends a basicMC by allowing supervisory feedback on its output Y from any part of the AI system even *after* the call. For example, this feedback may be for overriding the output Y to Z like in a supervisoryMC, except that a consumer of the output Y may provide this feedback at a later time. A reviewerMC may allow feedback other than overrides, for example, for rewards in reinforcement learning [13].

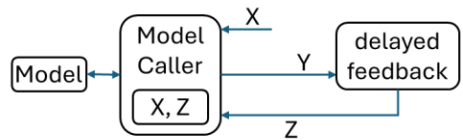

Figure 7: A reviewerMC allows delayed feedback from consumers of its output Y, for example, by overriding it with Z. When the model caller gets this feedback, it updates its data to replace Y by Z.

## 2.7 HostMC

A hostMC extends a basicMC so that it can be used as a surrogate of a function, that is, any call to the function gets automatically routed through the hostMC, as illustrated in Figure *8*. Technically, the actual name of the function is rebound to the hostMC.

A hostMC is useful in the AI transformation use case of Section 1.1.1. Suppose the legacy FDC (fraud-detection component) is exposed as a function *fdc* that takes transaction parameters as input and outputs a fraud score for the transaction based on some hand-coded rules. Now suppose that the AI team has developed a model *mfd* (model for fraud detection) that has the same input/output signature, and we want to upgrade FDC so that model *mfd* gets automatically called whenever the function *fdc* is called. This is easily achieved by creating a hostMC with model *mfd* and host *fdc* in a parallel aggregation.

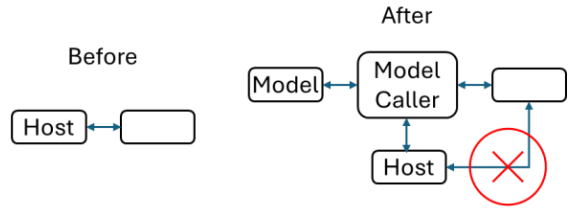

Figure 8: A hostMC becomes a surrogate of its host, so that all calls previously routed to the host (Before) are now routed to the hostMC (After). The hostMC may call the host to compute its output.

To be more useful, a hostMC allows a variety of behaviors controlled by its *call-target* attribute:
- If the call-target attribute is set to the string 'host', then only the host gets called, which in this case will be *fdc* and the system will behave as the legacy system, except that input/output data is saved in hostMC.
- If the call-target attribute is set to '*registered*', then only the registered model gets called, which in this case calls only model *mfd*. Thus, the system behaves as if *fdc* has been upgraded to *mfd*, as shown in Figure 8.
- If the call-target is set to '*both*' then both host and hostMC are called as if they are in a parallel ensemble and their individual outputs are aggregated to obtain the final output. Thus, the system becomes a hybrid of legacy and the updated system, exhibiting a novel behavior.

For AI transformation of any function *fdc*, we generally recommend the following steps:
A. Create a hostMC for function *fdc* with a registered model *mfd*
B. Set its call-target to 'host'.
C. Keep saving input/output data, while using it to fine-tune model *mfd*, and testing it to determine whether it matches the desired (or required) quality.[12]
D. If the model becomes qualified, then change the call-target to '*both*' so that both the model and the host are used.
E. Once the model has been fully validated (perhaps by meeting an even higher quality-threshold), then change the call-target to '*registered*', so that the function *fdc* is no longer used.
F. Remove function *fdc* whenever the system needs to be simplified.

For convenience, a hostMC may automate all the above steps, except for A and F.[13]

A host in a hostMC may be a model, instead of a function. Its behavior is similar, except that the hostMC is now a surrogate for the host model and the original model is called (when the call-target is set to 'host' or '*both*') instead of the original function.

## 2.8 FunctionMC

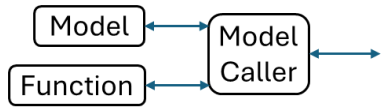

Figure 9: A functionMC allows functions to be also registered in a model caller, just like a model.

A functionMC extends a basicMC by adding a function directly in an ensemble with its model, as illustrated in Figure *9*. Thus, the final output of calling a functionMC is obtained by calling both the model and the function, similar to that in an ensembleMC with two models (think of a model there being replaced by a function here). However, functionMC does not become a surrogate for the function, as in a hostMC, since registering is different from hosting.

---

[12] The desired quality is specified in another attribute, called *quality-threshold*. A model that meets or exceeds the desired quality is considered *qualified*, otherwise it is considered *unqualified*.
[13] Our python library also provides a convenient function for easily executing Step A, as described later in Section 3.

Let's consider a simple version of functionMC that behaves similar to a parallelMC: both the function and the model are called in parallel, and their outputs are aggregated to obtain the final output. In this simple case, both the function and the model must have the same input and output parameters.

## 2.9 ContextMC

A contextMC extends a hostMC by automatically adding parameters from the program context of the host call, as illustrated in Figure *10*. While X indicates the parameters of the host function, parameters Y from the program context of the call are also added in the auto-modified call to contextMC, which becomes a surrogate of the host. This allows sending more data to the model, similar to what happens in an IdMC. Since the id of an idMC is part of the context, an idMC is a specific kind of contextMC.

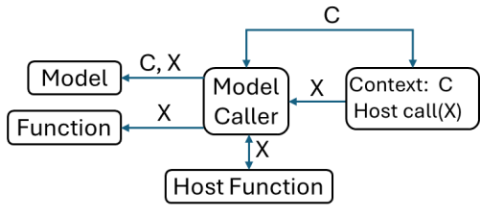

Figure 10: A contextMC retrieves context arguments C form the call context and sends it to its registered model along with the host function arguments X, while sending only X to its host and registered functions.

## 2.10 GatedMC

A gatedMC extends an ensembleMC by activating only a subset of models for any given inputs, similar to a sparse mixture of experts (MoE) model [14], as illustrated in Figure *11*. Only the last two models are activated here for a specific call. Different models may be activated for different inputs. Not activating all models reduces load on computational and network resources, which also leads to savings in energy, cost, and response latency.

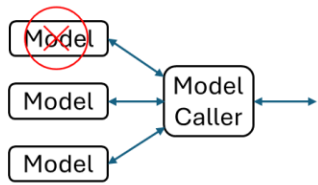

Figure 11: For any given call, a gatedMC activates a subset of registered models (the last two here), reducing the computational load and potentially leading to savings in energy, cost, and response latency.

## 2.11 CollaborativeMC

A collaborativeMC extends a basicMC by adding an interface for collaboration with people or other systems, as illustrated in Figure *12*. This interface is similar to that for a supervisoryMC, except that the output from this interface is used in a parallel ensemble for aggregation, instead of for overriding the aggregated output.

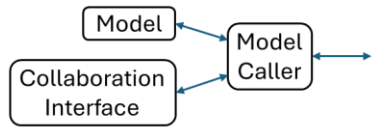

Figure 12: A collaborativeMC exposes an interface for collaborating with users or other systems, as if they were models or functions.

### 2.12 AnytimeMC

An anytimeMC extends a basicMC by generating a monotonic sequence of outputs with increasing expected quality, instead of a single output. This requires at least one more output source besides a model, which could be a host or some component in an ensemble. This also requires some estimation of output qualities (accuracy, etc.).

An anytimeMC provides the slow and fast thinking capability of System 2 and System 1, respectively [15], but in a single system. The fast outputs are of lower quality than the slower outputs of higher quality. Instead of just one fast and one slow output, an anytimeMC generates a sequence of multiple increasingly slower (that is, decreasingly faster) and better-quality outputs.

### 2.13 ProgrammableMC

A programmableMC extends a basicMC by allowing programming of some of its execution, instead of full configuration, as illustrated in Figure *13*. There are various ways of adding this custom code, including defining new subclasses and injecting code in the call. The custom code may contain some learnable parameters, such as an AI model. Thus, it may be possible to train and evaluate the programmableMC itself, in addition to the models registered with it. Some special kinds of ProgrammableMCs are described below. A programmableMC enables future-proofing needed for the use case of Section 1.1.9.

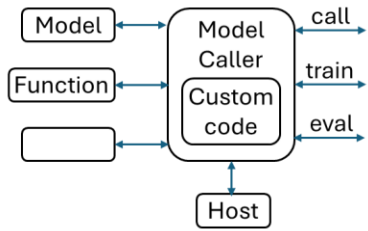

Figure 13: A programmableMC allows programming of custom code to exhibit more sophisticated behavior than available through just configurations. When the custom code contains learnable parameters then it may be trained and evaluated just like models.

#### 2.13.1 CustomAggregationMC

The behavior of an ensembleMC may be programmed, instead of using the limited configurations described in Section 2.3. For example, it may filter out all models below a specified expected accuracy and then take a mean of the outputs from the remaining ones.

#### 2.13.2 {Framework/Library}MC

The host and registered models and functions may be chained together using a specific framework or library such as langchain[14], llamaindex[15], DSPy [16], and OpenAI function calling[16] for external tools.

---

[14] https://github.com/langchain-ai/langchain
[15] https://www.llamaindex.ai/

For example, a DSPyMC may define a custom DSPy module for RAG (retrieval-augmented generation) using built-in modules such as dspy.Retrieve and dspy.ChainOfThought. It may then use the cached training data to compile and optimize the module using some DSPy teleprompter (optimizer), before using it for regular calls. The registered functions may access external tools, knowledge graphs, etc.

### 2.14 FullMC

A fullMC is a model caller that combines all the capabilities described above. It contains entities, including:
- Registered models and functions
- A host model or function
- Variety of datasets
- supervisory and collaborative interfaces
- Id and context injections
- Sensory and reviewer callbacks
- Model and function gating
- Anytime outputs
- Configurations
- Automations
- Custom code

## 3 A unifying framework for model callers

While Section 2 presented several kinds model callers, it raises an important question: can we provide a unifying framework for model callers that covers all those kinds? We present such a unification in this section and demonstrate its flexibility by creating even more kinds of model callers!

### 3.1 Basic concepts

While some of the basic concepts presented in this section may be familiar, we review their essential aspects and also mention some non-traditional aspects:

A *function* is any entity that may be called (or invoked) with some inputs (also called function arguments) to produce some outputs. These arguments are operational (i.e., runtime) values of parameters specified in the function definition. The signature of a function specifies the datatypes of its inputs and outputs. This notion of function is very general and includes methods in object-oriented programming languages. User interfaces and external APIs are also functions, where a user or an external system or service produces the outputs.

A *dataset* is a collection of input/output pairs of a function. It is a subset of the full (mathematical) relation defined by a (computational) function. A dataset may also be used to specify the desired (or required) behavior of a function.

A function may be evaluated using a dataset to assess its accuracy based on an exact or some approximate matching of actual versus desired outputs.[17] Technically, a function provides a method called *eval* which may be called with a dataset argument for its evaluation. Thus, calling a function is different from calling the eval method of the function.[18]

---

[16] https://platform.openai.com/docs/guides/function-calling
[17] This is slightly non-traditional since only models are traditionally evaluated!
[18] Some functions provide a method, such as *predict* or *forward*, for calling the function.

A *model* (instance) is any function that can be trained using a dataset, for the purpose of improving its accuracy.[19] Training modifies a model by updating values of its learnable parameters, often using an optimizer. We explicitly differentiate models from other functions since model callers are purposely designed to provide special support for them. Technically, a model provides a method called *train* which may be called with a dataset argument for its training. Training is also influenced by values assigned to meta-parameters like the learning rate of the optimizer.

### 3.2 Model callers

Conceptually, a model caller is an entity that encapsulates models, functions, and datasets to enable the kind of activities described in this section.

*Registering models and other functions with specific roles and attributes*: A model caller allows registering a function or a model with specific roles and specific attribute values. Roles allow a model caller to differentiate among its registered components, for example, using some registered models as members of a parallel ensemble while using a different registered model for aggregating the outputs from the ensemble members. The attributes may be used in many ways, such as assigning specific optimizers to a model and using a specific learning rate for it. Informally, functions and models registered in a model caller are available to it during operations.

*Surrogating functions and models with specific attributes*: A model caller surrogates a function (or a model) if any call to the function or its method is instead routed to the model caller. In other words, the model caller (or its method) gets called instead of the function or the model (or its method). A function surrogated by a model caller is called a *host* of the model caller. Informally, a model caller takes over the operations of its hosts. The attributes of a host may be used in many ways, for example, the value of the attribute *call-target* specifies the behavior when the host is called:
- '*host*': only the host is called
- '*registered*': some or all registered models and functions are called (depending on their roles and the inputs)
- '*both*': both the host and the registered models and functions are called

*Calling:* A model-caller may be called just like a function. If uses its registered components to generate the output, similar to calling a host whose *call-target* is set to '*registered*'.

*Storing and managing datasets:* A model caller enables storing datasets of various kinds, depending on how they were collected (say, with supervision or without) and their purpose (say, for training or evaluation). These might include gold, platinum, silver, and bronze kinds described in Section 2.2.

*Interacting using methods and callbacks*: Though both of these are functions, methods are part of the standard interface of a model caller, while callbacks are custom functions embedded in the outputs generated by the model caller (so only their recipients have access). Methods may include setting, retrieving, and updating attributes and parameter values, datasets, registering/unregistering, etc. Similar to a function, a model caller may also provide an "*eval*" method for evaluating its accuracy for some dataset.

*Configuring its behavior:* The operational behavior of a model caller is influenced by its configuration which is defined by values of specific attributes called configuration parameters. The behaviors may include automating actions, such as caching operational inputs and outputs in datasets, or imposing constraints, for example, allowing at most one host.

*Programming its behavior:* A model caller may include custom code to influence its operational behavior. There are various ways of adding this custom code, including defining new subclasses of model callers and injecting code while calling a model caller. The custom code may contain some learnable parameters, such as an AI model.

*Learning of parameter values*: Some model caller attributes may be learnable, similar to those in models. Similarly, such model callers also provide a "*train*" method for training using some dataset.

### 3.3 New kinds of model callers

The unifying framework of model callers presented above enables creating some new kinds of model callers and some new behaviors that were not explicitly described in Section 2.

#### 3.3.1 Novel constraints

- A model caller might require that only those registered functions and models may be activated whose evaluation accuracy on its golden (and/or silver) dataset exceeds a specified threshold.

#### 3.3.2 Novel behaviors

- A model caller may provide multiple training behaviors:
    - Local training: update trainable parameters of the model caller only
    - Nested training: also train models registered with an "ensemble" role
    - Fresh training: reinitialize learnable parameters before training
    - Incremental training: continue training without reinitializing learnable parameters

- A model caller may provide multiple evaluation behaviors, such as:
    - Golden evaluation: evaluate using only supervised evaluation datasets
    - Combined evaluation: evaluate using all evaluation datasets
    - Local and nested evaluation: similar to the corresponding training behaviors

#### 3.3.3 Sharing among model callers

- Multiple model callers may register the same model or function: Even if the same component is registered in multiple model callers, its evaluation may differ since it's influenced by the local dataset saved in each of those model callers. In contrast, updating a shared model by training it using the local dataset of one model caller will also update it for all other model callers.

#### 3.3.4 Other behaviors

- When hosted, an original function or model may get hidden (since its name is reused by the model caller) or may remain exposed (by giving a new name to the model caller).
- Model callers may be *nested*, i.e., a model caller may register or surrogate other model callers.
- Model callers may enforce user-role-based access-control such that only some methods (facades) are available to specific roles. For example, a SWE may be allowed to add hosts, only a ML engineer may be allowed to add models.
- Model callers may provide a convenient target or home for managing or governing AI systems.

## 4 `ModelCaller` python library

We have implemented an initial partial prototype of model callers as a python class `ModelCaller` and have released it as a pip-installable python library called `modelcaller`, which is also available for download on GitHub. The GitHub site also contains a jupyter notebook with an introductory session on how to use this library. We present just a high-level description in this section, especially the aspects not covered in earlier sections.

A `ModelCaller` object may be created in one of the following ways:
- Wrapping a predefined function or model by calling the `mc_wrap` function. This creates a new `ModelCaller` object (instance of the `Modelcaller` class) whose attribute `host` is set to the original unwrapped function or model. This function returns a surrogate for the host, which may be called instead of the host to provide the model-caller capabilities.
- Decorating the definition of a function or a model by `@mc_wrapd` decorator. This uses the `mc_wrap` function to create a new `ModelCaller` instance whose attribute `host` is set to the original unwrapped function or model. This function also returns a surrogate for the host, which may be called instead of the host to provide the model-caller capabilities.
- Creating an instance of `ModelCaller` in the standard way. In this case, the `host` is initialized to `None`.

The core components of a `ModelCaller` object are as follows:
- `models`: a list of AI models that are registered through this object. All these models must have the same input/output signatures.
- `functions`: a list of functions in this objects ensemble which also includes the models in `models`. All these functions must have the same input/output signatures. Further, the output parameters must be identical to those of the models and the input parameters must be a subset of those of the models. The model parameters that are not function parameters are called *context* parameters, since they are automatically obtained from the programmatic context of the function calls.
- `edata` and `tdata`: a variety of labeled datasets that may be used for training and evaluating the models. Some of these datasets may be *supervised*, that is, the labels have been created or validated by experts, while the others are *semi-supervised*, that is, the labels have been created by some system and not yet validated by experts.

A ModelCaller object also has the following key attributes that control its behavior, including some automations:
- `host`: the original model or a function that was wrapped to create this ModelCaller object (`None` otherwise). It is not required that the `host` be a member of either the `functions` or the `models` list.
- `auto_cache`: if set to `True` then all the inputs and outputs passing through the object during operations are automatically saved to `io_data`.
- `edata_fraction`: the fraction of auto-cached data saved for evaluation, instead of training.
- `feedback_fraction`: the fraction of aggregated outputs to be validated by a supervisor before being confirmed.
- `quality_thresholds`: a pair of numbers that specify the supervised and unsupervised accuracies, respectively, that must be met by a model before its outputs are used in the ensemble. A model is considered *qualified* if and only if (iff) it meets or exceeds both of these thresholds.
- `auto_train`: if set to `True` then any model added to the `ModelCaller` object is automatically trained using io_data.
- `auto_test`: if set to `True` then any model is automatically tested for accuracy after its training.

- `auto_id`: if set to `True` then the `ModelCaller` object id is added as a context argument to each model call.[20]
- `call_target`: specifies the targets of any call to the host's surrogate. If this is set to the string 'host' then the original unwrapped host is called. If this is set to the string "`registered`" then only the function and models in the two lists are called, similar to directly calling the ModelCaller object.[21] If it is set to 'both' then both the original host and the ModelCaller objects are called. While the results are currently aggregated in a fixed way, future versions of ModelCaller will allow customizing the aggregation logic.

# 5    Other related work

We now discuss additional related work that was not explicitly mentioned earlier.

Several tools automatically cache datasets of input/output pairs such as databricks inference tables[22]. Many existing tools encapsulate datasets with models, such as pipeline with memory in scikit-learn[23] and trainer class in HuggingFace transformers library[24]. However, none of these make them available as a first-class programming language object that provides the full functionality of a model, while enabling training and evaluating using those datasets. However, they are designed for other purposes and provide capabilities not provided by model callers.

Besides the ones mentioned in Section 2.13.2, many other frameworks have been proposed to simplify the development, serving and management of generative AI systems. Some of these have overlapping capabilities, such as FrugalGPT [17] that allow efficient routing through LLM ensembles that overlaps with GatedMC of Section 2.10.

# 6    Conclusions

We presented a new SW abstraction, called model callers, for building and managing AI systems, several practical use cases, many kinds of model callers, a framework for unifying them all, and a python library with a prototype implementation. Model callers essentially encapsulate AI/ML models, SW 1.0 functions, and datasets in a single first-class object in a programming language.

We believe AI systems of the future should be built using this higher-level abstraction of model callers, instead of the currently used lower-level abstraction of models. This is similar to the transition from using functions to using objects in object-oriented systems that completely changed software engineering (SWE) [18], [19].

Model callers also allow building AI systems using the best SWE practices that have been perfected over several decades, instead of staring with data and models, as is the current practice. The SWE team can build systems out of model callers, just like they do out of traditional objects and functions; the ML/AI/DS teams can then refine those model callers to provide the desired performance. We believe that ML/AI models have matured enough to make this transition now.

Figure *14* illustrates a pictorial meta-summary of this paper. (A) represents the current situation in which AI and SW (non-AI) domains are deeply intertwined leading to mess and inefficiencies (for

---

[20] Same happens when `auto_id` is set to `None`, except that all the other capabilities of the `ModelCaller` object are disabled, making it a very thin passthrough object.

[21] With one subtle difference: calling the ModelCaller object directly requires model arguments (including context arguments, that is, the values of the context parameters), while the surrogate (with `call_target` set to '`registered`') may be called with just the function arguments (without context arguments).

[22] https://www.databricks.com/blog/announcing-inference-tables-simplified-monitoring-and-diagnostics-ai-models

[23] https://scikit-learn.org/stable/modules/generated/sklearn.pipeline.Pipeline.html

[24] https://huggingface.co/docs/transformers/main_classes/trainer

example, SW engineers need to become AI experts and AI engineers need to become SW experts). (B) represents our aspiration to clean up the current situation by decoupling the two domains using an appropriate intermediary. This raises a big question: what is the right intermediary? (C) provides one answer that "models" are the right intermediaries between the AI and SW domains, and that the AI domain "owns" the models and model servers. This is the dominant view today, mainly because this question has not been asked and no better intermediaries have been proposed, to the best of our knowledge. (D) provides our answer that "model callers" are the right intermediaries between the AI and SW domains, and that the SW domain should own them. Model callers allow plugging in of models (via registration and model servers) which should continue to be owned by the AI domain.[25] We believe that this strikes the right balance between the two domains while allowing them to collaborate efficiently with each other and with other domains.[26]

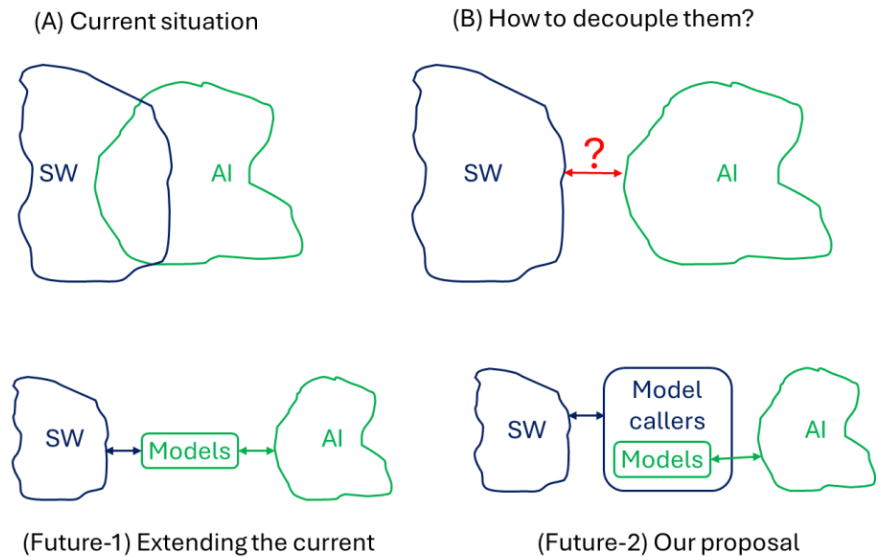

Figure 14: (A) Current interaction between the AI and SW (non-AI) domains: tight coupling leading to confusion and inefficiencies. (B) A big question: how to decouple them? (C) One potential future that simply extends the current: connect the two domains using models and serving systems owned by the AI domain (D) Our proposed future: connect the two domains using model callers (owned by the SW domain) that contains models (owned by the AI domain)

---

[25] This does not prevent the AI domain in owning model callers that are eventually (after possibly a few nesting) plugged as models into the SW-owned model callers.

[26] We can also borrow a metaphor from science fiction that has also been used for web applications: model callers are "portals" in the SW dimension (or universe) that contain models or "portlets" into the AI dimension!